# High-speed domain wall racetracks in a magnetic insulator

Saül Vélez[1,§,*], Jakob Schaab[1,§], Martin S. Wörnle[1,2], Marvin Müller[1], Elzbieta Gradauskaite[1], Pol Welter[2], Cameron Gutgsell[1], Corneliu Nistor[1], Christian L. Degen[2,*], Morgan Trassin[1,*], Manfred Fiebig[1] and Pietro Gambardella[1,*]

[1] Department of Materials, ETH Zurich, 8093 Zurich, Switzerland

[2] Department of Physics, ETH Zurich, 8093 Zurich, Switzerland

[§] These authors contributed equally

[*] e-mail: saul.velez@mat.ethz.ch (S.V.); degenc@ethz.ch (C.L.D.); morgan.trassin@mat.ethz.ch (M.T.); pietro.gambardella@mat.ethz.ch (P.G.)

**Recent reports of current-induced switching of ferrimagnetic oxides coupled to a heavy metal layer have opened realistic prospects for implementing magnetic insulators into electrically addressable spintronic devices. However, key aspects such as the configuration and dynamics of magnetic domain walls driven by electrical currents in insulating oxides remain unexplored. Here, we investigate the internal structure of the domain walls in $Tm_3Fe_5O_{12}$ (TmIG) and TmIG/Pt bilayers and demonstrate their efficient manipulation by spin-orbit torques with velocities of up to 400 m s$^{-1}$ and minimal current threshold for domain wall flow of 5 x 10$^6$ A cm$^{-2}$. Domain wall racetracks embedded in TmIG are defined by the deposition of Pt current lines, which allow us to control the domain propagation and magnetization switching in selected regions of an extended magnetic layer. Scanning nitrogen-vacancy magnetometry reveals that the domain walls of thin TmIG films are Néel walls with left-handed chirality, with the domain wall magnetization rotating towards an intermediate Néel-Bloch configuration upon deposition of Pt. These results indicate the presence of a sizable interfacial Dzyaloshinskii–Moriya interaction in TmIG, which leads to novel possibilities to control the formation of chiral spin textures in magnetic insulators. Ultimately, domain wall racetracks provide an efficient scheme to pattern the magnetic landscape of TmIG in a fast and reversible way.**

Spintronics relies on the use of current-induced torques for manipulating the magnetization of thin films and nanodevices (*1*). Owing to spin-orbit coupling, charge currents flowing in heavy metals, such as Pt, Ta or W, generate spin currents that exert a torque onto an adjacent ferromagnetic layer (*2*, *3*). These so-called spin-orbit torques (SOTs) are capable of reversing the magnetization of ferromagnets in a highly efficient and ultra-fast manner (*4–8*) as well as driving domain walls at very high velocities (*9–11*). Most studies in this area, however, have been performed on ultrathin metallic ferromagnets, for which extensive characterizations of the domain wall (DW) structure and velocity have been reported (*12–20*).

Magnetic insulators offer exciting perspectives for spintronic and magnonic applications beyond conventional metallic systems (*21*). In particular, ferrimagnetic rare-earth garnets coupled to heavy metal layers have attracted attention due to the possibility of electrically exciting and detecting propagating magnons (*22–25*) as well as for their low power and high frequency magnetization dynamics (*26*). Despite increasing interest in such systems, however, the electrical manipulation of the equilibrium magnetization has not been investigated in detail. Current-induced switching of magnetic insulators has been only recently demonstrated in $Tm_3Fe_5O_{12}$ (TmIG) and $BaFe_{12}O_{19}$ in combination with either Pt or W layers (*27–30*). These studies relied on magnetoresistance measurements in order to detect the orientation of the magnetization, from which the dynamics of the switching process cannot be inferred. In this work, we present a combined scanning nitrogen-vacancy (NV) magnetometry and spatially-resolved magneto-optic Kerr effect (MOKE) study of the DW structure and dynamics driven by SOTs in racetrack structures embedded in a TmIG layer. We demonstrate highly efficient current-induced DW motion in TmIG/Pt, with mobility comparable or larger than metallic ferromagnets, an extraordinarily low threshold for DW flow, and very small depinning fields. We further provide the first direct characterization of the DW width and internal structure in thin-film TmIG as well as TmIG/Pt bilayers. Previous studies in garnets were only able to provide estimates of the DW width based on indirect or diffraction-limited optical measurements, with reported values ranging from tens of nanometers to micrometers (*31–34*). Scanning NV magnetometry reveals that the DWs in TmIG films are only ~20 nm wide and have a well-defined chiral structure, which changes from left Néel in TmIG to intermediate left Néel-Bloch in TmIG/Pt. These findings evidence the presence of strong interfacial Dzyaloshinskii–Moriya interaction (DMI) in TmIG grown on substituted gadolinium gallium garnet $Gd_3Sc_2Ga_3O_{12}$ (SGGG), which is attenuated by the deposition of Pt. The DMI is the key ingredient required to stabilize chiral Néel DWs in ferromagnets and ferrimagnets with perpendicular magnetization, which can then be driven by SOTs at very high velocities (*10–15*). In contrast with metallic ferromagnets, TmIG thin films support the formation of Néel DWs without introducing heavy metal layers. Our results show that ferrimagnetic garnets are ideal materials for fabricating efficient and high-speed DW racetracks.

TmIG(8 nm)/Pt(5 nm) bilayers were grown on SGGG (111)-oriented substrates by a combination of pulsed laser deposition for epitaxial growth of the garnet and in-situ dc sputtering for Pt. The numbers between parentheses indicate the thickness of each layer. Pt current lines were patterned in the shape of Hall bars by optical lithography and selective etching of the metal, leaving the TmIG film un-etched (see Methods). Figure 1A shows an optical image of a TmIG/Pt device. The structural, electric, and magnetic characterization of the TmIG film and the TmIG/Pt bilayer are reported in the Supplementary Materials (Section S1 and Figs. S1 and S2).

The magnetic state of the TmIG film underneath the Pt current line, +**m** (up) or -**m** (down), can be read electrically by measuring the transverse Hall resistance $R_{xy}$, as shown in Fig. 1B during a sweep of the out-of-plane magnetic field $H_z$. The measurement confirms that the films exhibit robust perpendicular magnetic anisotropy with a coercive field of about 40 Oe. In agreement with a previous report (*27*), the magnetization of TmIG can be deterministically switched upon the application of a current pulse of sufficient current density $J_x$ in the presence of a constant in-plane field $H_x$ (Fig. 1C). The switching polarity is determined by the damping-like component of the SOT (*2–4*), which stabilizes +**m** for $J_x$ parallel to $H_x$ and –**m** for $J_x$ antiparallel to $H_x$ in TmIG/Pt. Notably, we found that full switching can be achieved with pulses of 1 ms at current densities below $10^7$ A cm$^{-2}$ for an in-plane field as small as $|H_x|$=20 Oe, which confirms the high quality of our devices (see also Supplementary Materials Section S1).

A distinctive feature of our experiments is that TmIG covers the entire substrate, but switching occurs only in the region defined by the Pt current line. This is clearly seen in Fig. 1D, which shows a differential MOKE image of a TmIG/Pt device after the application of a current pulse. The bright contrast coinciding with the Pt current line shows the region where the magnetization has switched from -**m** to +**m**, demonstrating that it is possible to control the magnetization of a continuous TmIG film in a local way without altering the magnetic moments of the surroundings. Only in the presence of an out-of-plane field, or in the case of significant Joule heating, the switched magnetic domain may extend beyond the Pt line (see the Supplementary Materials Section S2). Such local control of the magnetization is unique to magnetic insulators due to the confinement of the current in the metal overlayer. As discussed further below, we also find that the surrounding magnetic medium influences the switching dynamics underneath the Pt current line. This is seen by the fact that a larger current is required for inducing down-to-up switching relative to up-to-down switching when starting from a homogenously magnetized TmIG film pointing down (Fig. 1C), and vice versa starting from a homogenously magnetized layer pointing up (not shown).

As SOT-induced switching is strongly dependent on the DW structure (*12–15*), we use scanning NV magnetometry to reveal the DW magnetization profile in both TmIG and TmIG/Pt layers. The technique is based on a single NV defect located at the apex of a diamond tip, which senses the magnetic stray field $B_{NV}(X,Y)$ emanating from a magnetic surface with high spatial resolution (Fig. 2A) (*17*, *35*, *36*). Figure 2B shows $B_{NV}(X,Y)$ of the TmIG film measured in a region where a DW intersects an area partially covered by Pt. From this measurement, we reconstruct the out-of-plane component of the magnetization $M_z(X,Y)$ (see Methods), as shown in Fig. 2C. Although the DW runs continuously across the Pt edge, the linescans of $B_{NV}(X,Y)$ shown in Figs. 2D and 2E reveal that the DW structure changes going from TmIG/Pt to TmIG. In order to extract the magnetization profile of the DW from these measurements, we fit $B_{NV}(X,Y)$ by assuming that the magnetization components in the rotated coordinate system XYZ (Fig. 2A) vary as (*17*, *33*)

$$M_X(X) = M_S \frac{\cos\psi}{\cosh\left(\frac{X}{\Delta_{DW}}\right)},$$

$$M_Y(X) = 0,$$

$$M_Z(X) = -M_S \tanh\left(\frac{X}{\Delta_{DW}}\right), \quad (1)$$

where $\psi$ defines the angle of the in-plane magnetization direction with respect to the $X$-axis, $\Delta_{DW}$ is the DW width, and $M_S$ the saturation magnetization. Figures 2D and 2E compare representative $B_{NV}(X,Y)$ line profiles for TmIG/Pt and TmIG together with fits performed assuming a pure Bloch wall ($\psi = 90°$), left Néel wall ($\psi = 180°$) and right Néel wall ($\psi = 0°$). The best fit of the stray field map in Fig. 2B according to Eq. 1 gives $\Delta_{DW} = (17 \pm 17)$ nm and $\Delta_{DW} = (27 \pm 6)$ nm for TmIG/Pt and TmIG, respectively (see Methods). Despite the large uncertainty in $\Delta_{DW}$, which is due to the weak dependence of $B_{NV}$ on $\Delta_{DW}$, the fits show that the DWs in 8-nm-thick TmIG are extremely narrow. Moreover, we find that the DW have a Néel chiral structure in both TmIG/Pt and TmIG, which is characterized by the angles $\psi = (116 \pm 33)°$ and $\psi = (173 \pm 17)°$, respectively. Thus, the DW changes from left-handed Néel to an intermediate left-handed Néel-Bloch configuration in going from TmIG to TmIG/Pt, which is a compelling indication of the presence of negative DMI in the bare TmIG layer, most likely due to symmetry breaking at the SGGG/TmIG interface. The deposition of Pt reduces the DMI, which we ascribe to the presence of positive DMI at the TmIG/Pt interface, consistently with the sign of the DMI found in metallic ferromagnetic/Pt bilayers (*14*, *15*, *17*). These findings have important consequences for the operation of DW racetracks in magnetic insulators, because the reduced $\Delta_{DW}$ favors the localization of DWs, whereas the finite DMI allows for their efficient manipulation by SOTs.

In order to prove this last point, we investigate the switching dynamics and current-induced DW motion below the Pt line. We refer to the switching of the magnetization starting from a homogeneously magnetized TmIG layer as *forward switching* (*domain expansion*), and to the return to a homogenous magnetic state starting from a reversed domain as *backward switching* (*domain contraction*). Figure 3A shows the relative change in the magnetization induced by a single forward switching current pulse as a function of $H_x$ and $J_x$. Within the experimental error, we find that the switching diagram is symmetric upon inversion of $H_x$ or $J_x$, indicating that the SOT efficiency is the same for up-to-down and down-to-up switching and independent on the current direction. Distinct to electrical reading (*27–29*), which is only sensitive to the magnetic moments in the vicinity of the Hall cross (Fig. 1, B and C, and Fig. S5), MOKE measurements reveal that for a wide range of $H_x$ and $J_x$ only partial switching is achieved. We thus study the distribution and evolution of reversed magnetic domains induced by a sequence of current pulses. Since we expect an influence of the surrounding TmIG on the magnetization dynamics underneath the Pt current line (Fig. 1C), and because the switching process is symmetric upon inverting both **m** and **H** (*4*, *8*), we investigate the forward and backward switching processes for one fixed initial state of the film (-**m**).

Figures 3B and 3C show two representative sequences of differential MOKE images taken during forward switching and backward switching, respectively. For each case, we compare the combinations of field and current that allow for domain nucleation and expansion ($\pm H_x, \pm J_x$) and domain contraction ($\pm H_x, \mp J_x$). These images reveal that forward switching occurs via nucleation of a reversed domain at a defect site (as confirmed by a series of repetitions) and subsequent domain expansion along the Pt current line, with comparable speeds for both DWs on the left- and right-hand sides of the domain. Backward switching takes place by pushing the outer DWs towards the center of the domain. Similar dynamics –for either expansion or contraction– is observed upon inverting $H_x$ and $J_x$. The different timescales of the switching processes (see Figs. 3B and 3C) indicate that domain contraction is significantly faster than domain expansion. The most likely cause for this behavior is the tendency of the reversed domain to shrink in order to reduce the DW energy (*37*). Accordingly, we find that the minimum pulse length required to induce DW motion upon contraction is much smaller than for expansion (Fig. 3D).

Measurements of the DW velocity $v_{DW}$ are reported in Fig. 4 for an up-down DW as a function of $J_x$ and $H_x$ during both domain expansion and domain contraction. $v_{DW}$ is evaluated by considering the total DW displacement after a sequence of current pulses of length $t_p$ as the DW moves along the Pt current line. Since the same $t_p$ does not allow for sampling a large $H_x, J_x$ parameter space, we used longer (shorter) pulses for smaller (larger) $H_x, J_x$ values. Note that the DW velocity remains almost constant when changing $t_p$, indicating that it is not influenced by inertia (see Methods and Fig. S6). Our measurements reveal robust DW velocities of up to ~200 m s$^{-1}$ for domain expansion (Fig. 4A) and ~400 m s$^{-1}$ for domain contraction (Fig. 4B), which are comparable to the ones found in all-metallic structures under similar conditions (*9*, *14*, *15*). Most remarkably, however, the DW mobility $\mu_{DW} =$ $\frac{v_{DW}}{J_x}$ reaches values in excess of $3 \times 10^{-10}$ m³ A⁻¹ s⁻¹ for $J_x = 5 \times 10^7$ A cm⁻², which is comparable to that observed in compensated metallic ferrimagnets (*11*, *38*). In contrast, most metallic ferromagnets feature $\mu_{DW} = 0$ in this current range (*9*, *10*, *14*, *15*).

The linear increase of $v_{DW}$ with $J_x$ for expanding and contracting walls (Fig. 4, A and B) further reveals a very low onset of the DW flow regime ($\lesssim 5 \times 10^6$ A cm⁻²) compared to conventional ferromagnetic layers (*9*, *10*, *14*, *15*). This behavior is attributed to the reduced depinning field of TmIG, ~1-2 Oe (see Fig. S7), which is one to two orders of magnitude smaller than in metallic and semiconducting ferromagnets with perpendicular magnetic anisotropy (*39*). Upon increasing the current, $v_{DW}$ reaches a plateau between ~0.5 x 10⁸ A cm⁻² and ~1.0 x 10⁸ A cm⁻², followed by a further upturn. The plateau indicates the saturation of the DW velocity at $v_{DW}^{sat} \approx \gamma \Delta_{DW} \frac{\pi}{2}(H_x + H_{DMI})$, which occurs at a current density $J_x \gg \frac{2e\alpha\mu_0 M_s t}{\hbar \theta_{SH}}(H_x + H_{DMI})$, where $\gamma$ is the gyromagnetic ratio, $\hbar$ the reduced Planck constant, $\mu_0$ the vacuum permeability, $\theta_{SH}$ the effective spin Hall angle of Pt, and $H_{DMI}$ the effective DMI field (*12*, *16*). Taking $\gamma \sim 1.43 \times 10^7$ Oe⁻¹ s⁻¹ (*40*) and $\Delta_{DW} \sim 20$ nm (Fig. 2), and by approximating $H_x + H_{DMI} \approx H_x = 300$ Oe, we find that $v_{DW}^{sat} \sim 135$ m s⁻¹, which agrees well with the experimental data (Fig. 4, A and B). The further increase of $v_{DW}$ beyond saturation, which is typical also of metallic ferromagnets (*10*, *15*), is attributed to the influence of Joule heating and the Oersted field (see the Supplementary Materials Section S2 and Figs. S4 and S8).

In agreement with the presence of DMI inferred from the DW magnetization profile, we observe a slightly larger $v_{DW}$ when the DW moves against the direction of the current (red symbols in Fig. 4). The same behavior is also confirmed for down-up DWs (Fig. S9). This asymmetry, which is characteristic of chiral Néel DWs (*14*, *15*), is consistent with the left-handed Néel chirality derived from scanning NV magnetometry and the sign of the torques in TmIG/Pt (see Fig. S10 for more details). By fitting the $v_{DW}(H_x)$ data to a linear function and extrapolating it to $v_{DW} = 0$ (*15*), we estimate $|H_{DMI}| \sim 12 \pm 3$ Oe (Fig. 4, C and D). The DMI constant can then be calculated as $D = \mu_0 H_{DMI} M_S \Delta_{DW}$ (*9*, *14*), where $M_S \sim (6.0 \pm 1.5)$ x 10⁴ A m⁻¹ and $\Delta_{DW} \sim 20$ nm, yielding $D \sim -2 \pm 2$ µJ m⁻². This value is two to three orders of magnitude smaller than the DMI reported for ultrathin metallic ferromagnets/Pt (*14*, *15*, *18*) and one to two orders of magnitude smaller than that of ferrimagnetic metal/Pt bilayers (*11*). As the DW mobility in the flow regime is proportional to $\Delta_{DW}/\alpha M_s$, the large mobility and high DW velocities in TmIG/Pt appear as the direct consequence of the small $M_s$ and low-damping $\alpha$ typical of garnet layers (*31*, *41*). By tuning the interfacial DMI, we anticipate that even larger $v_{DW}$ may be reached at a relatively low current density.

Our results demonstrate fast current-driven DW motion in a magnetic insulator and provide first insight into the internal DW structure of thin garnet layers. The chiral Néel structure of the DWs in TmIG reveals that oxide interfaces support a finite DMI even in the absence of heavy metal layers, which makes it possible, in principle, to stabilize nontrivial topological configurations in insulating magnetic thin films, such as spin spirals and skyrmions. The low current threshold for DW flow and the large DW mobility, combined with the possibility of defining DW racetracks embedded in a continuous magnetic medium, make TmIG extremely attractive for spintronic applications. Local control of the magnetization is unique to magnetic insulators, which opens the possibility of printing arbitrary circuit paths enabling, for instance, the implementation and *in situ* reconfiguration of synthetic magnetic structures with tailored magnonic bands (*21*, *42*) and nanomagnonic waveguides (*43*, *44*).

## Methods:

**Films growth and devices fabrication.** The TmIG thin films were grown by pulsed laser deposition on (111)-oriented $Gd_3Sc_2Ga_3O_{12}$ substrates (lattice constant $a$ = 12.56 Å) to achieve high tensile strain (~2%), which promotes perpendicular magnetic anisotropy (*45*). The substrate temperature was 650 °C, the oxygen pressure was 0.2 mbar, while the laser fluence and repetition rate were set to 1.35 J cm$^{-2}$ and 8 Hz, respectively. After deposition, the samples were cooled in 200 mbar oxygen at a rate of -10 K/min. The bare TmIG films showed a root-mean-square roughness of about 0.15 nm over a ~5 x 5 µm$^2$ area. After growth, the samples were directly transferred to the sputter chamber without breaking vacuum, where the Pt layer was deposited at room temperature for three minutes at a power of 10 W in 0.05 mbar Ar. The thickness of the layers were calibrated by x-ray reflectometry. For the sample presented in the main text, the thicknesses of TmIG and Pt where 8.3 and 5.0 nm, respectively. The Pt layer was patterned into Hall bars (consisting of three Hall crosses separated by $L = 50$ µm with a total channel length of 140 µm and width $W = 10$ µm) by photolithography and subsequent Argon plasma etching.

**Electric transport measurements.** The longitudinal and transverse Hall resistances $R_{xx} = V_x/I_x$ and $R_{xy} = V_y/I_x$, respectively, were measured by applying an alternating current (AC) of amplitude $I_x = 0.3$ mA and frequency $f = 11$ Hz and by recording the first harmonic longitudinal ($V_x$) and transverse ($V_y$) voltages, as shown schematically in Fig. 1a.

**MOKE measurements.** We used a home-built wide-field polar MOKE microscope with Koehler illumination to measure the out-of-plane component of TmIG. Magnetic contrast was enhanced by taking differential MOKE images, i.e., each image was subtracted by a reference image captured in a fully magnetized state. The setup was equipped with two sets of orthogonal coils for the generation of out-of-plane and in-plane magnetic fields. For the switching and DW velocity studies, current pulses were injected using an AGILENT 8114A (100V/2A) pulse generator with a 50 Ω output impedance. The impedance matching with the Pt current line was achieved by connecting a 50 Ω resistance in parallel to the current line.

The relative change in the magnetization shown in Fig. 3A was evaluated by integrating the differential MOKE signal along the Pt current line (corresponding to the bright area in Fig. 1D) after the application of a single current pulse starting from a fully magnetized state.

For the domain expansion measurements (Fig. 4, A and C), an initial domain was nucleated by a single current pulse at a defect site near the center of the Hall bar. For the domain contraction experiments (Fig. 4, B and D), the initial domain was generated by switching the area underneath the Pt current line with a single current pulse of $t_p$ = 150 ns, $|J_x| = 0.94 \times 10^8$ A cm$^{-2}$ and $|H_x|$ = 125 Oe, leading to a domain as the one shown in Fig. 1D.

In order to compare the DW velocities obtained for domain expansion and domain contraction, we studied the same DW moving back and forth over the same area. The case presented in Fig. 4 corresponds to an up-down DW moving between the center and the right end of the Hall bar. The DW velocity $v_{\mathrm{DW}}$ was evaluated by measuring the total DW displacement $\Delta x$ (as identified by direct MOKE imaging) obtained after the application of a series of $N_\mathrm{p}$ current pulses of width $t_\mathrm{p}$, yielding $v_{\mathrm{DW}} = \Delta x / (N_\mathrm{p} \, t_\mathrm{p})$. The pulses were applied at a frequency of 1 Hz in order to minimize the heat load during the experiment. $v_{\mathrm{DW}}$ was found to be nearly independent of $t_\mathrm{p}$ –only showing a slight increase of 10% or less when doubling the pulse length, which we attribute to Joule heating– indicating that the DW motion coincides with the pulse duration (see Fig. S6 for more details).

**Scanning NV magnetometry.** Spatially resolved scans of the magnetic stray field produced by a DW in TmIG (see Fig. 2A) were acquired on a home-built nanoscale scanning diamond magnetometer (NSDM) microscope. Experiments were carried out in ambient environment and at zero magnetic bias field. The NSDM employed a monolithic diamond probe tip with a single NV center implanted at the apex (QZabre LLC, www.qzabre.com). The NV center spin resonance was monitored by optically-detected magnetic resonance (ODMR) spectroscopy (*36*)(*46*) using a nearby microwave antenna (~2.9 GHz) for spin excitation and fluorescence microscopy (532 nm excitation, 630-800 nm detection) for spin state readout. To convert the spin resonance frequencies to units of magnetic field, we fitted the ODMR spectrum to a double Lorentzian and extracted the frequency difference $\Delta f$ between the resonance peaks. The detected field $B_{NV}$ is then given by $|B_{NV}| = \frac{\pi \Delta f}{\gamma}$, where $\gamma = 2\pi \cdot 28.0$ GHz/T is the electron gyromagnetic ratio. To re-establish the relative sign of $B_{NV}$, we inverted ($B_{NV} \rightarrow -B_{NV}$) the image on one side of the DW (Fig. 2B). Note that scanning NV magnetometry provides a vector projection of the magnetic field,

$$B_{NV} = \vec{B} \cdot \vec{e}_{\mathrm{NV}} = \sin\theta_{NV} \cos\phi_{NV} B_x + \sin\theta_{NV} \sin\phi_{NV} B_y + \cos\theta_{NV} B_z, \qquad (2)$$

because the NV center is sensitive only to fields that are parallel to its symmetry axis $\vec{e}_{\mathrm{NV}}$. Here, $\vec{B}$ is the vector field at the position of the NV center, and $\theta_{NV}$ and $\phi_{NV}$ are the polar and azimuth angles of $\vec{e}_{\mathrm{NV}}$ in the laboratory frame (see Fig. S11). The direction $\vec{e}_{\mathrm{NV}}$ is determined by the crystallographic orientation of the diamond tip and the probe arrangement in the setup (see Fig. 2A). $\theta_{NV}$ and $\phi_{NV}$ were calibrated by a series of ODMR measurements and confirmed by line scans. For the experiments presented in Fig. 2, $\theta_{NV} = (55 \pm 2)°$ and $\phi_{NV} = (83 \pm 3)°$.

We investigated the magnetization, spin structure and width of the DW by analyzing the local field image $B_{NV}(X,Y)$ shown in Fig. 2B. In a first step, we fitted line cuts across the TmIG to TmIG/Pt step edge to extract the NV center stand-off distance, $d = (104 \pm 5)$ nm (see the Supplementary Materials Section S3). To characterize the chirality and width of the DW, we took line cuts of $B_{NV}$ perpendicular to the DW as shown in Fig. 2B, and compared them to the analytical model given through Eq. (1). The associated magnetic stray field was obtained by forward propagation of Eq. (1) in $k$-space according to (*47*)(*48*),

$$\hat{B}_X = \hat{g}(-k\hat{M}_X + ik_X\hat{M}_Z),$$

$$\hat{B}_Y = \hat{g}(-k\hat{M}_Y + ik_y\hat{M}_Z),$$

$$\hat{B}_Z = \hat{g}(ik_X\hat{M}_X + ik_Y\hat{M}_Y + k\hat{M}_Z), \tag{3}$$

where hat symbols indicate Fourier transforms, $k_X$, $k_Y$, and $k = (k_X^2 + k_Y^2)^{1/2}$ are the in-plane $k$-space vectors, $\hat{g} = \frac{\mu_0 t}{2}\left(\frac{1-e^{-kt}}{kt}\right)e^{-kd}$ is the Fourier transform of the Green's function (*48*), and $t = 8.3$ nm is the film thickness. When taking line cuts along $X$ across a domain wall extending along $Y$, the $Y$ and $\hat{M}_Y$ terms become zero and Eq. (3) simplifies to $\hat{B}_X = \hat{g}(-k\hat{M}_X + ik_X\hat{M}_Z)$, $\hat{B}_Y = 0$ and $\hat{B}_Z = \hat{g}(ik_X\hat{M}_X + k\hat{M}_Z)$.

To extract values for $M_s$, $\psi$ and $\Delta_{DW}$ we fitted the experimentally measured $B_{NV}$ to the analytical prediction by Eqs. (1-3), with the domain wall position $x = x_0$ as an additional fit parameter. By repeating the fitting procedure for a series of line scans, we obtained distributions for all parameters together with their means and standard deviations. A detailed description of the fitting procedure and error analysis is given in the Supplementary Materials Section S3.

By assuming that the magnetization is predominantly out-of-plane, we further reconstructed a magnetization map $M_z$ from the magnetic field map $B_{NV}$ by using

$$\hat{M}_Z = \frac{-\hat{B}_{NV} W}{g(ik_X e_X + ik_Y e_Y - k e_Z)}, \tag{4}$$

where $W$ is a low-pass filter (cutoff $\lambda = d$) that suppressed high spatial frequencies in the image (*48*). The resultant magnetization map is plotted in Fig. 2C. Note that while the magnetic domains of TmIG and TmIG/Pt are well reproduced, the reconstruction slightly overestimates $M_z$ near the DW due to the left Néel character of the DW.

## Supplementary Materials

Supplementary material for this article is available at .........

- Section S1. Electrical characterization of TmIG/Pt by spin Hall magnetoresistance: determination of the interface quality and the magnetic anisotropy
- Section S2. Influence of the presence of an applied out-of-plane field, current-induced Oersted fields and Joule heating on the magnetization switching process
- Section S3. Scanning NV magnetometry
- Fig. S1. Structural and magnetic characterization of the thulium iron garnet thin films
- Fig. S2. Electrical characterization of the spin Hall magnetoresistance and magnetic anisotropy of TmIG/Pt
- Fig. S3. Forward switching in the presence of an out-of-plane field $H_z$
- Fig. S4. Switching dynamics at large current densities: influence of the Oersted fields
- Fig. S5. Simultaneous Hall and MOKE measurements of TmIG/Pt devices
- Fig. S6. Dependence of $v_{DW}$ on the pulse length
- Fig. S7. Evaluation of the domain wall depinning field
- Fig. S8. Multiple domain nucleation at large current densities
- Fig. S9. Comparison of $v_{DW}$ between an up-down and a down-up domain wall
- Fig. S10. Schematics of the domain wall structure and current-induced domain wall motion in the presence of $H_x$ and DMI
- Fig. S11. Definition of the polar angles $\phi_{NV}$ and $\theta_{NV}$
- Fig. S12. In situ calibration of the NV stand-off distance on TmIG/Pt
- References (*50–55*)

## Acknowledgements

This work was supported by ETH Zurich, the Swiss Competence Centre for Materials Science and Technology (CCMX), by the Swiss National Science Foundation under Grants No. 200020-172775 and by the European Research Council through the Advanced Grant No. 694955—INSEETO. The authors thank Rudolf Schäfer and Eva Grimaldi for discussions, and Kevin Chang and Jan Rhensius for help in construction of the scanning NV magnetometer as well as for providing diamond tips. M.F. thanks ETH Zurich and CEMS at RIKEN for support of his research sabbatical. C.L.D. acknowledges funding by the Swiss National Science Foundation under Grant No. 200020-175600 and the NCCR QSIT, and by the European Commission through Grant No. 820394 "ASTERIQS".

## Author contributions

S.V., J.S., M.T., and P.G. conceived the study. J.S., E.G., and M.T. grew and characterized the magnetic and structural properties of TmIG and TmIG/Pt. S.V. fabricated the TmIG/Pt devices and performed and analyzed the electrical measurements. S.V., J.S., and M.M. performed the MOKE measurements; S.V. and J.S. analyzed the data. M.M., C.G., and C.N. built the wide-field MOKE setup. M.S.W., P.W., and C.L.D. built the scanning NV magnetometer and performed and analyzed the scanning NV measurements. M.T., M.F., C.L.D., and P.G. supervised the work. S.V. and P.G. wrote the manuscript. All authors contributed to the scientific discussion and manuscript revision.

## Competing interests

The authors declare that they have no competing interests.

## Data and materials availability

All data needed to evaluate the conclusions in the paper are present in the paper and/or the Supplementary Materials. Additional data related to this paper may be requested from the authors.

# Figures

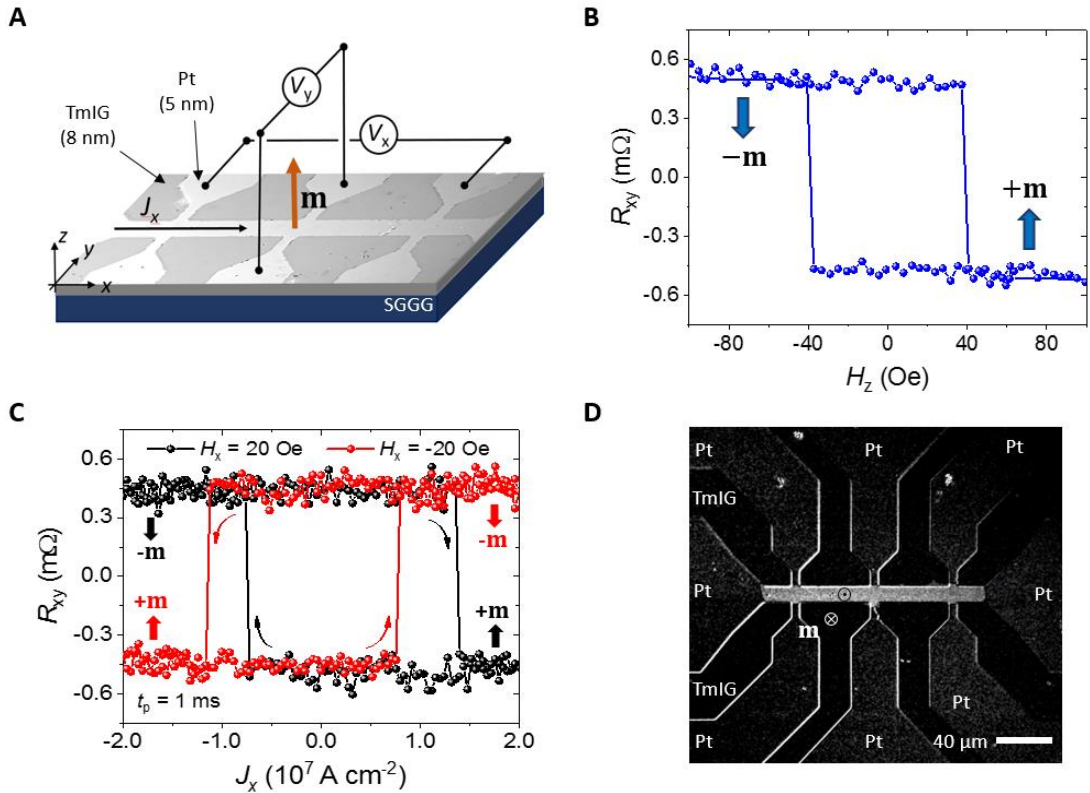

**Fig. 1. Device schematics and local switching of TmIG.** (**A**) Optical image of a Pt Hall bar patterned on TmIG with superposed electric wiring, coordinate system, and magnetization vector **m**. (**B**) Hall resistance $R_{xy}$ as a function of $H_z$. The anomalous Hall-like signal arises from the interaction of the spin current generated in the Pt layer with the out-of-plane magnetization component $m_z$ of TmIG (*27, 49*), leading to a high (low) $R_{xy}$ for -**m** (+**m**). The data are shown after subtraction of a constant sample-dependent offset. (**C**) Electrical measurement of current-induced switching of TmIG ($t_p$ = 150 ns, $H_x$ = + 20, black dots, and -20 Oe, red dots). Note that, starting from a film saturated in the –**m** state, higher current densities are required to induce forward switching relative to backward switching. The same behavior is observed for $\pm H_x$, thus ruling out a misalignment of the sample as a possible explanation for this effect. (**D**) Wide-field differential MOKE image of a TmIG/Pt device after injection of a current pulse ($J_x$ = 0.94 x $10^8$ A cm$^{-2}$, $t_p$ = 150 ns, $H_x$ = 100 Oe). The film was initially saturated in the –**m** state by applying a magnetic field $H_z$ = -100 Oe. The bright contrast shows that only the TmIG region underneath the Pt current line has switched to +**m**.

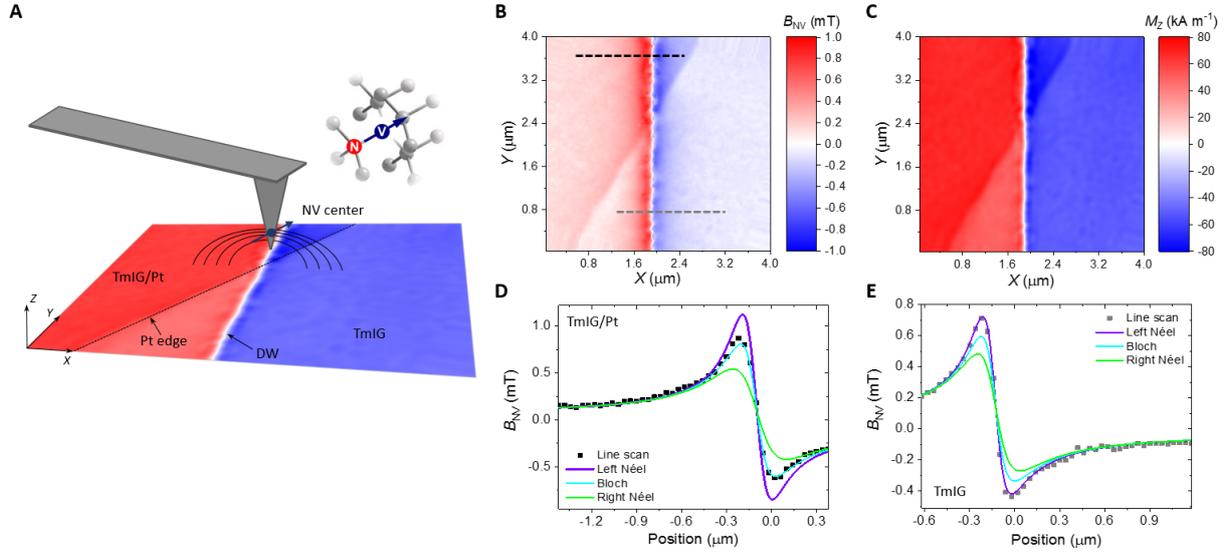

**Fig. 2. Domain wall structure and chirality in TmIG and TmIG/Pt measured by scanning NV magnetometry.** (**A**) Schematic of the NV probe and sample. The color code represents the out-of-plane component of the surface magnetization of the film, with red (blue) corresponding to areas of opposite magnetization. A DW crosses the TmIG/Pt and TmIG regions. The inset shows a NV center within the diamond lattice and the corresponding spin quantization axis. (**B**) Stray field $B_{NV}(X,Y)$ measured by scanning the diamond tip over the sample surface shown in (A). (**C**) Reconstructed out-of-plane magnetic map $M_z(X,Y)$ from the data shown in (B) (see Methods). The larger $M_z = (74.6 \pm 1.7)$ kA m$^{-1}$ of TmIG/Pt relative to $M_z = (44.1 \pm 2.5)$ kA m$^{-1}$ of TmIG is mainly attributed to the proximity-induced polarization of the Pt layer (deduced from measurements performed on a reference TmIG layer, see the Supplementary Materials Section S3). (**D** and **E**) Line scans of $B_{NV}$ along the dashed lines indicated in (B) (square dots). The solid lines are fits of $B_{NV}$ profiles assuming pure Bloch (cyan), left-handed Néel (violet) and right-handed Néel (green) DW structures.

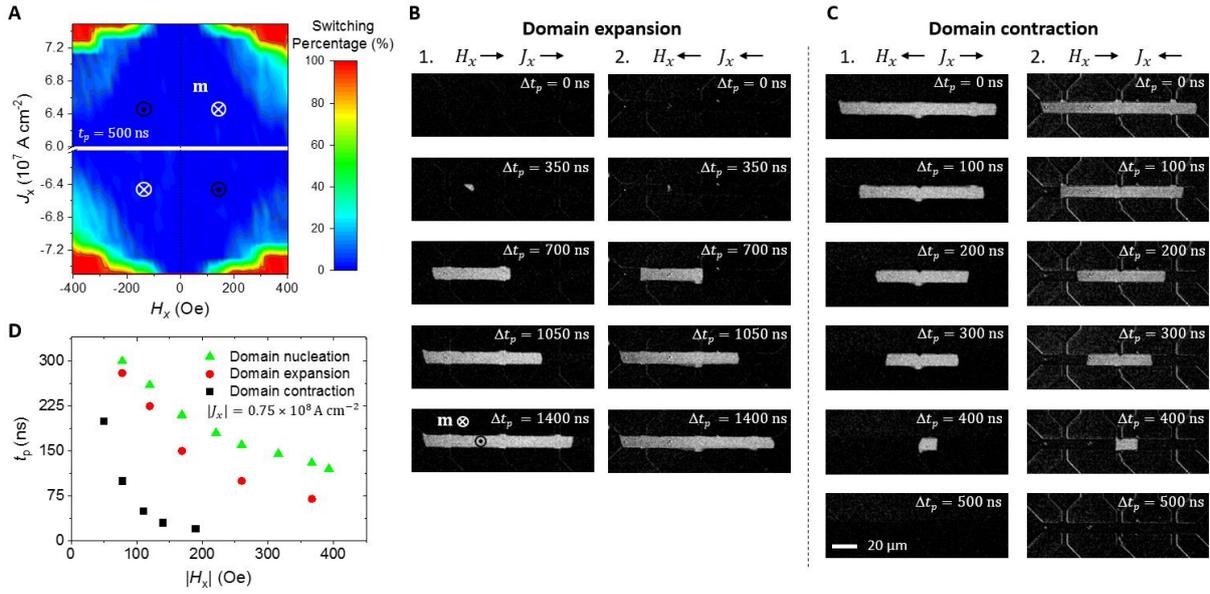

**Fig. 3. Current-induced DW dynamics in embedded TmIG racetracks.** (**A**) Magnetization switching diagram showing the percent change of the magnetization underneath the Pt current line as a function of $J_x$ and $H_x$. For each switching event, the initial **m** was always set fully up or down by applying an out-of-plane field $H_z$ = +100 or -100 Oe, respectively (+**m** for $\pm H_x$, $\mp J_x$ and −**m** for $\pm H_z$, $\pm J_z$ according to the symmetry of SOT switching). Forward switching is triggered by a single current pulse with $t_p$ = 500 ns. (**B**) Sequence of differential MOKE images showing current-induced domain nucleation and expansion and (**C**) domain contraction for the different combinations of $H_x$ and $J_x$ that allow each scenario to take place. $\Delta t_p = N t_p$ is the accumulated pulse time after applying $N$ current pulses of length $t_p$. The pulse length is $t_p$ = 350 and 50 ns and the current density $|J_x|$ = 0.65 x 10$^8$ and 0.75 x 10$^8$ A cm$^{-2}$ for (B) and (C), respectively. $|H_x|$ = 250 Oe in all cases. Bright and dark MOKE contrasts correspond to +**m** and −**m** states, respectively. Note that the TmIG film not covered by Pt remains fully down magnetized. (**D**) Switching diagram showing the minimum pulse length $t_p$ required to nucleate a domain starting from a fully saturated film (green triangles), to expand a domain (red circles), and to contract a domain (black squares) as a function of $|H_x|$. The current density of the pulses is fixed to $|J_x|$ = 0.75 x 10$^8$ A cm$^{-2}$. The results show that it is easier to move a DW than to nucleate a domain, and that the onset of DW motion is lower for contraction relative to expansion.

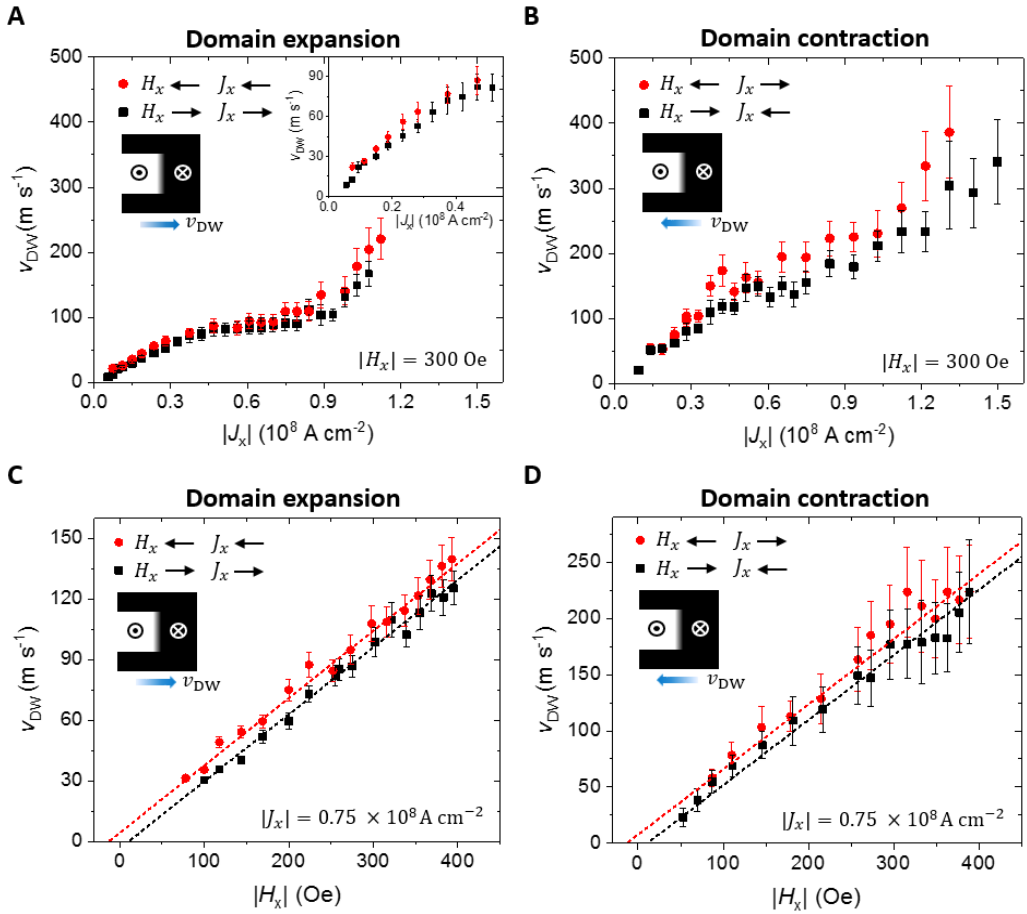

**Fig. 4. DW velocity upon expansion and contraction of domains in TmIG/Pt surrounded by uniformly magnetized TmIG.** (**A**) $v_{DW}$ of an up-down DW (corresponding to a +**m** domain in a –**m** medium) as a function of $J_x$ upon expansion and (**B**) contraction. The magnetic field is set to $|H_x|$ = 300 Oe. The inset in (A) shows an enlarged view of $v_{DW}$ in the flow regime at low current density. (**C**) $v_{DW}$ of an up-down DW as a function of $H_x$ upon expansion and (**D**) contraction. The current density is fixed at $|J_x| = 0.75$ x $10^8$ A cm$^{-2}$ in either case. DWs move faster when $J_x$ is opposite to $v_{DW}$ (red symbols), indicating the presence of a weak DMI favoring left-handed Néel chiral DWs. Dashed lines in either (C) and (D) are linear fits to the experimental data assuming same slope of $v_{DW}(J_x)$ for both polarities of $J_x$ and opposite magnetic field at which $v_{DW}$ crosses zero, which allows evaluating the effective DMI field $|H_{DMI}| \sim 12 \pm 3$ Oe (*15*).

# Supplementary Materials for

## High-speed domain wall racetracks in a magnetic insulator

Saül Vélez[1,§,*], Jakob Schaab[1,§], Martin S. Wörnle[1,2], Marvin Müller[1], Elzbieta Gradauskaite[1],
Pol Welter[2], Cameron Gutgsell[1], Corneliu Nistor[1], Christian L. Degen[2,*], Morgan Trassin[1,*],
Manfred Fiebig[1] and Pietro Gambardella[1,*]

[1] Department of Materials, ETH Zurich, 8093 Zurich, Switzerland

[2] Department of Physics, ETH Zurich, 8093 Zurich, Switzerland

[§] These authors contributed equally

[*]e-mail: saul.velez@mat.ethz.ch (S.V.); degenc@ethz.ch (C.L.D.); morgan.trassin@mat.ethz.ch (M.T.);
pietro.gambardella@mat.ethz.ch (P.G.)

**This pdf file includes:**

- Section S1. Electrical characterization of TmIG/Pt by spin Hall magnetoresistance: determination of the interface quality and the magnetic anisotropy
- Section S2. Influence of the presence of an applied out-of-plane field, current-induced Oersted fields and Joule heating on the magnetization switching process
- Section S3. Scanning NV magnetometry
- Fig. S1. Structural and magnetic characterization of the thulium iron garnet thin films
- Fig. S2. Electrical characterization of the spin Hall magnetoresistance and magnetic anisotropy of TmIG/Pt
- Fig. S3. Forward switching in the presence of an out-of-plane field $H_z$
- Fig. S4. Switching dynamics at large current densities: influence of the Oersted fields
- Fig. S5. Simultaneous Hall and MOKE measurements of TmIG/Pt devices
- Fig. S6. Dependence of $v_{DW}$ on the pulse length
- Fig. S7. Evaluation of the domain wall depinning field
- Fig. S8. Multiple domain nucleation at large current densities
- Fig. S9. Comparison of $v_{DW}$ between an up-down and a down-up domain wall
- Fig. S10. Schematics of the domain wall structure and current-induced domain wall motion in the presence of $H_x$ and DMI
- Fig. S11. Definition of the polar angles $\phi_{NV}$ and $\theta_{NV}$
- Fig. S12. *In situ* calibration of the NV stand-off distance on TmIG/Pt
- References (*50–55*)

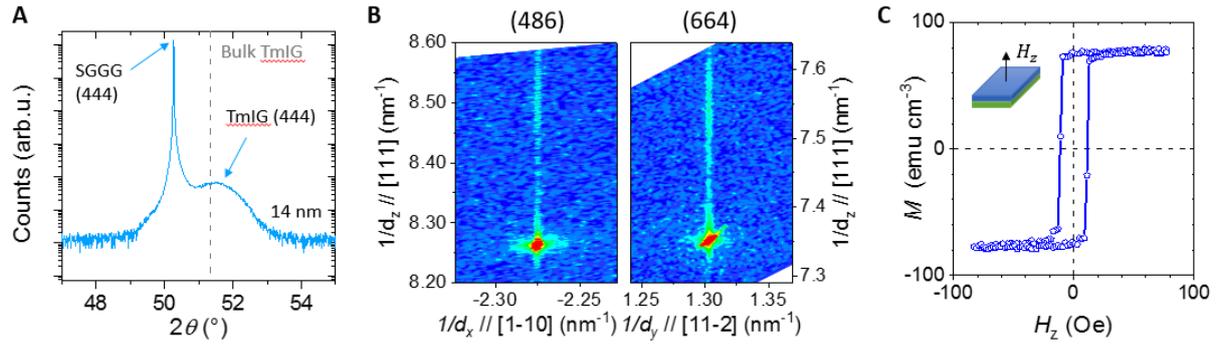

**Fig. S1. Structural and magnetic characterization of the thulium iron garnet thin films.** (**A**) Symmetric X-ray diffraction scan around the (444) peak of a SGGG/TmIG(14 nm) sample. The TmIG (444) peak shifts towards higher angles because of a reduction of the out-of-plane lattice constant of the film provided by the tensile strain (gray dashed line indicates the (444) peak position of bulk TmIG, extracted from Ref. (*50*)). (**B**) Reciprocal space maps of the same sample around the (486) and (664) substrate peaks. The coinciding in-plane lattice constants of film and substrate along the two perpendicular directions [1-10, 11-2] confirm full epitaxy. The same strain state is expected for the thinner TmIG(8.3 nm)/Pt sample discussed in the main text as strain is maintained for thicknesses up to ~50 nm (*45*). (**C**) Out-of-plane VSM measurement of the TmIG(8.3 nm)/Pt sample. A clear square-like hysteresis loop with a coercivity of about 15 Oe is observed, confirming that the film exhibits perpendicular mangetic anisotropy. The determined saturation magnetization $M_s$ (~75-80 emu cm$^{-3}$) is slightly reduced compared to bulk TmIG (110 emu cm$^{-3}$, (*31*)), possibly attributed to a slight oxygen deficiency caused by the low oxygen pressure during growth (*45*).

## Section S1. Electrical characterization of TmIG/Pt by spin Hall magnetoresistance: determination of the interface quality and the magnetic anisotropy

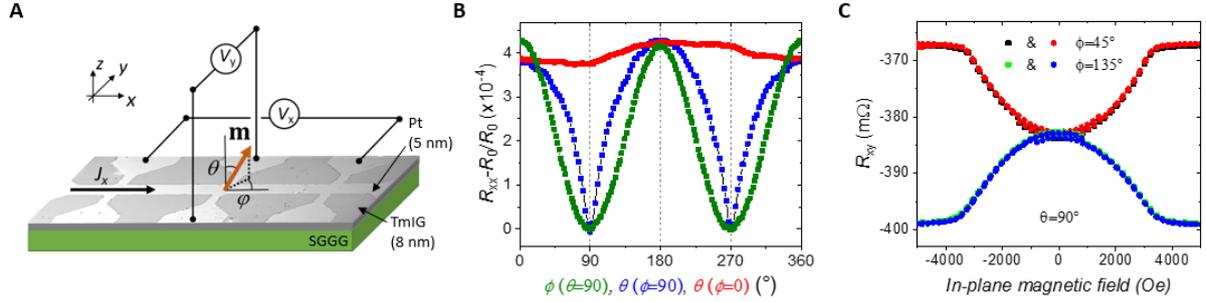

**Fig. S2. Electrical characterization of the spin Hall magnetoresistance and magnetic anisotropy of TmIG/Pt.** (**A**) Optical image of a patterned Pt Hall bar on the TmIG(8.3 nm) film presented in the manuscript with superposed electric wiring, coordinate system and the definition of the polar angles $\theta$, $\phi$, and the magnetization vector **m**. (**B**) Magnetoresistance measurements performed for **m** rotating in the *xy*- (green, $\theta = 90°$ and variable $\phi$), *yz*- (blue, fixed $\phi = 90°$ and variable $\theta$), and *xz*-plane (red, $\phi = 0°$ and variable $\theta$). The direction of **m** is controlled by rotating the sample in a magnetic field of amplitude 5 kOe. The angular dependences display the symmetry expected for spin Hall magnetoresistance (*49*). (**C**) Hall resistance $R_{xy}$ measurements performed by sweeping the magnetic field in the plane of the sample ($\theta = 90°$) at $\phi = 45°$ (black and red dots, trace and retrace) and $\phi = 135°$ (green and blue dots, trace and retrace). The variation of $R_{xy}$ with the applied magnetic field **H** is due to the canting of **m** –from the out-of-plane easy axis– towards **H**. The saturation of $R_{xy}$ at large magnetic fields indicates that **m** is saturated in the plane of the film, thus giving an estimation for the magnetic anisotropy field of the TmIG film of about $H_k \sim 3$ kOe.

According to the theory of the spin Hall magnetoresistance (SMR), the first harmonic response of the magnetoresistance in TmIG/Pt can be described as (*27, 49*):

$$R_{xx} = \frac{V_x}{I_x} = R_0 + \Delta R^{SMR}(1 - \sin^2\theta \sin^2\varphi), \qquad (S1)$$

$$R_{xy} = \frac{V_y}{I_x} = \frac{W}{2L}\frac{\Delta R^{SMR}}{2}\sin^2\theta \sin 2\varphi + R_H^{AHE,SMR}\cos\theta + R_H^{OHE}H_z, \qquad (S2)$$

where $\theta$ and $\varphi$ denote the polar and azimuthal angle of the normalized magnetization vector $\mathbf{m} = M/M_s(x,y,z)$ collinear to the applied field **H** (see Fig. S2A), W/2L = 0.1 is the width/length ratio of the measuring configuration of the Hall bar (i.e., 10/100, see Fig. S2A), $R_0$ is the longitudinal base resistance, $\Delta R^{SMR}$ the SMR amplitude, $R_H^{AHE,SMR}$ the anomalous Hall-like SMR amplitude, and $R_H^{OHE}H_z$ the ordinary Hall resistance. Fig. 1B shows the normalized $(R_{xx} - R_0)/R_0$ angular dependent magnetoresistance measurements performed in our sample with **m** rotating along the three main plains of the sample. $R_0$ is determined to be ~640 Ω. The change in resistance follows the expected behavior for SMR (*49*): i) a $\sin^2\theta$ dependence for **m** rotating in the plane of the film (green, $\theta = 90°$), ii) a distorted $\sin^2\theta$ dependence for **m** rotating in the *yz*-plane (blue, $\phi = 90°$) due to the strong demagnetizing field, and iii) no significant change when **m** rotates in the *xz*-plane (red, $\theta$ = 90°). The SMR amplitude $\Delta R^{SMR}/R_0 \approx 4.1 \times 10^{-4}$ agrees with previously reported values in TmIG/Pt structures (*27–29*). From the Hall measurements shown in Fig. S2C and Fig. 1B we estimate the $R_H^{AHE,SMR}$ amplitude to be ~-0.47 mΩ and the ratio $\frac{R_H^{AHE,SMR}}{R_0}\frac{2L}{W} \sim -0.734 \times 10^{-5}$, which is also consistent with earlier reports in TmIG/Pt (*27–29, 51*). From these measurements, we can estimate the real and imaginary parts of the spin-mixing conductance $G^{\uparrow\downarrow} = G_r + iG_i$ using the following expressions (*49*):

$$\frac{\Delta R_{SMR}}{R_0} = \theta_{SH}^2 \frac{\lambda_N}{d_N} \operatorname{Re}\left[\frac{2\lambda_N\ G^{\uparrow\downarrow}\ \tanh^2\frac{d_N}{2\lambda_N}}{\sigma_N + 2\lambda_N\ G^{\uparrow\downarrow} \coth\frac{d_N}{\lambda_N}}\right], \tag{S3}$$

$$\frac{R_H^{AHE,SMR}}{R_0}\frac{2L}{W} = -\theta_{SH}^2 \frac{\lambda_N}{d_N} \operatorname{Im}\left[\frac{2\lambda_N\ G^{\uparrow\downarrow}\ \tanh^2\frac{d_N}{2\lambda_N}}{\sigma_N + 2\lambda_N\ G^{\uparrow\downarrow} \coth\frac{d_N}{\lambda_N}}\right], \tag{S4}$$

where $\theta_{SH}$ is the spin-Hall angle, $\lambda_N$ the spin diffusion length, $d_N = 5.0$ nm the thickness and $\sigma_N = 3.13 \times 10^6$ Ω m the conductivity of the Pt layer. From the conductivity we estimate that $\lambda_N \sim 2.0$ nm and $\theta_{SH} \sim 0.049$ (*52*). By considering that $G_r \gg G_i$ (*53*), Eqs. (S3) and (S4) yield $G_r \sim 1.1 \times 10^{15}$ Ω$^{-1}$m$^{-2}$ and $G_i \sim 3.4 \times 10^{13}$ Ω$^{-1}$m$^{-2}$, with a corresponding ratio $G_r/G_i \sim 32$, which is consistent with literature values for magnetic insulator/heavy metal bilayers (*27, 28, 54, 55*) and confirms the high quality of the TmIG/Pt interface.

**Section S2. Influence of the presence of an applied out-of-plane field, current-induced Oersted fields and Joule heating on the magnetization switching process.**

As presented in the manuscript, a precise control of the magnetic moment underneath the Pt current line can be achieved in a continuous TmIG film by applying current pulses in the Pt line without altering the magnetic moments of the surrounding (Fig. 1D). However, the extremely small depinning field of the DWs in TmIG, ~1-2 Oe (see Fig. S7), makes the dynamics of the magnetization switching by current pulses to be extremely sensitive to the presence of any tiny out-of-plane magnetic field. Fig. S3 shows a sequence of differential MOKE images taken during forward switching in the presence of an out-of-plane field $H_z \sim 2$ Oe, showing that in contrast to what is observed in Fig. 1D and Fig. 3B, the switched domain clearly extends away from the Pt current line. The generation of out-of-plane Oersted fields at large current densities ($J_x \sim 10^8$ A cm$^{-2}$ or larger) also favors that the switched domain slightly extends away from the edges of the Pt current line (for about a couple of μm). This is seen in Fig. S3 (see image #7), where the switched domain extends more on the bottom side at the edge of the Pt current line, where the Oersted field favors the forward switching. The Oersted field has an even stronger influence on the switching dynamics during backward switching, as seen in Fig. S4. Finally, Joule heating produced by long current pulses ($t_p \sim 1$ μs or larger) at large current densities can also favor the switched domain to extend away from the Pt current line.

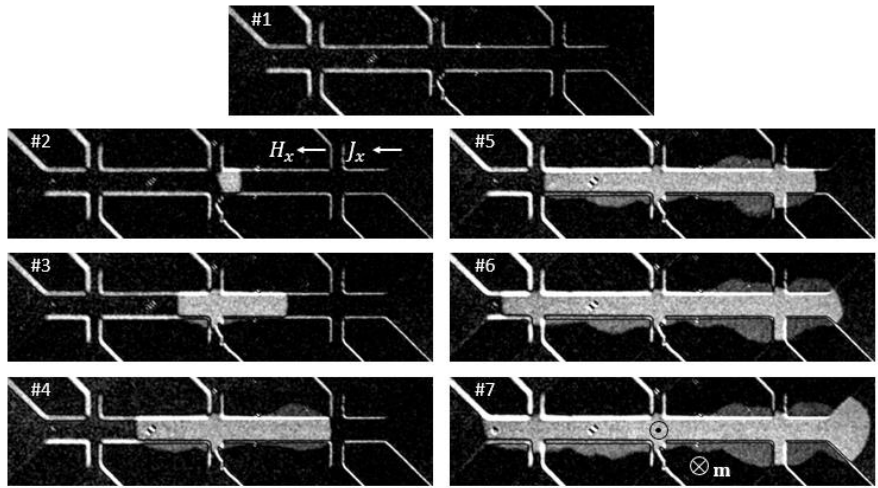

**Fig. S3. Forward switching in the presence of an out-of-plane field $H_z$.** Sequence of differential MOKE images taken during a series of forward switching current pulses of $J_x = 0.9 \times 10^8$ A cm$^{-2}$ and $t_p = 75$ ns in the presence of an external field of 400 Oe applied along the Pt current line at an angle of ~0.3° with respect to the plane of the film, which results in an out-of-plane field $H_z \sim 2$ Oe. The film is initially saturated down (-**m**, dark contrast). Bright contrast indicates the TmIG region that has switched to +**m**. The finite $H_z$ favors the spreading of the switched domain away from the current line. These measurements indicate that the accuracy in the alignment of the sample for the measurements presented in the manuscript is better than 0.1°.

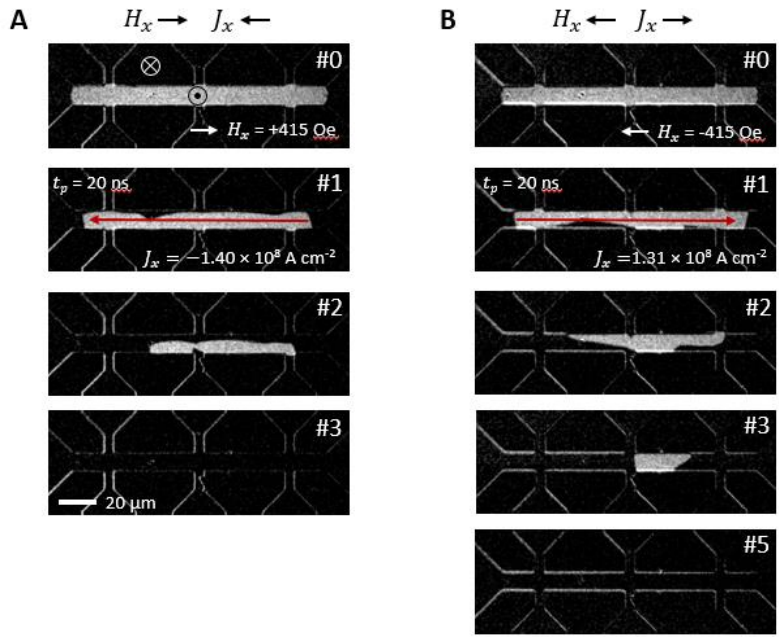

**Fig. S4. Switching dynamics at large current densities: influence of the Oersted field.** (**A** and **B**) Different examples of the dynamics of the DWs upon the application of a series of backward switching current pulses at large current densities. The current pulses produce out-of-plane Oersted fields of opposite polarity at the opposite edges of the Pt current line (i.e., bottom and top edges in the images). For a negative current, a positive (negative) out-of-plane Oersted field is generated at the bottom (top) edge of the Pt current line, and vice versa upon inverting the current direction. These Oersted fields can influence the switching dynamics given the extremely small DW depinning field in TmIG ~1-2 Oe (see Fig. S7). Therefore, for current pulses inducing the contraction of a +**m** domain (bright regions) in a -**m** medium (black regions), the negative out-of-plane Oersted field generated at the top (bottom) edge of the Pt current line for a negative (positive) current is sufficient to induce the depinning of the DW at the edge, pushing it towards the center of the line. As a result, the domain contraction has both transverse and longitudinal components, driven by the SOT and Oersted field, respectively (see images in A and B).

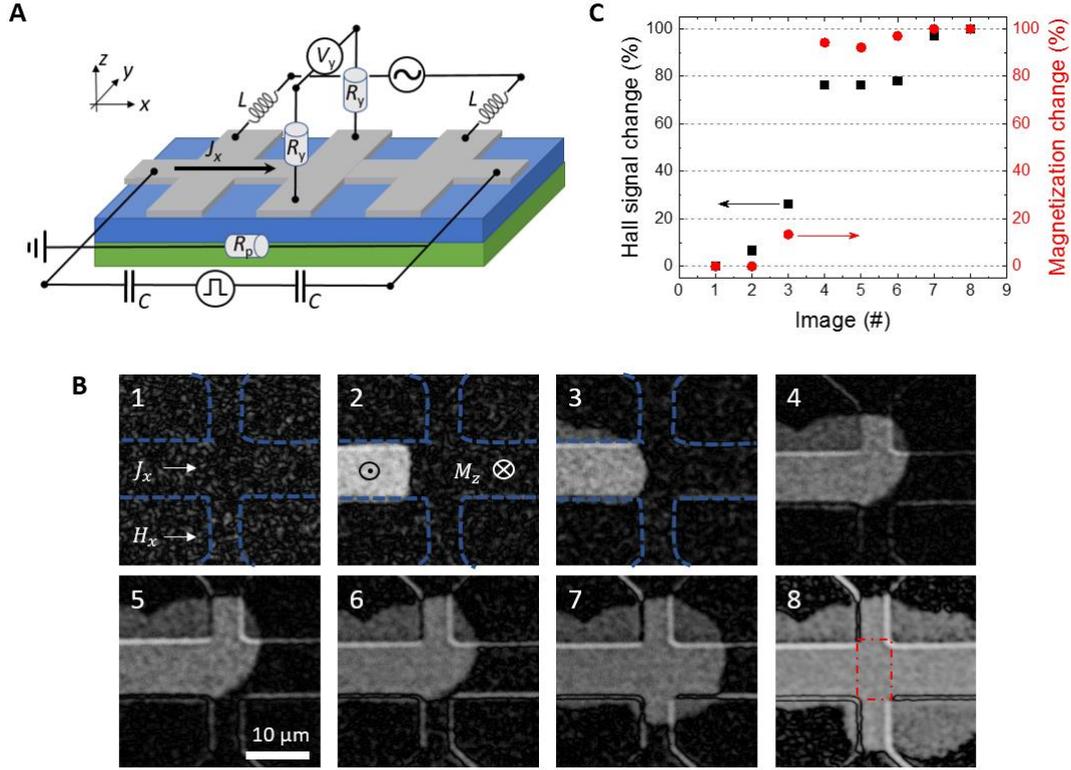

**Fig. S5. Simultaneous Hall and MOKE measurements of TmIG/Pt devices. (A)** Measurement scheme. Current pulses are applied via two capacitors (*C* = 150 nF) to the Hall bar, which allows current pulses faster than ∼ 1 µs to completely pass through while filtering low-frecuency signals. The current line is shunted by a 50 Ohm resistance $R_p$ to ensure optimum impedance matching. The AC current source for the harmonic Hall measurements (*f* = 11 Hz) is connected via inductors (*L* = 440 µH), which filter signals faster than ∼ 1 µs. The voltage probes for $V_y$ are connected through two resistors $R_y$ = 100 kΩ, which prevent damage of the Hall contacts from the current pulses. **(B)** Tracking of the current-induced expansion of a +**m** domain (bright area) in a down magnetized medium (-**m**, black). Between each image, a current pulse of $J_x$ = 7.5 x 10$^7$ A cm$^{-2}$ and adjustable pulse length (50 to 500 ns) is applied to slowly drive the DW through the Hall cross located at the center of the Hall bar (see A). The magnetic field is set at $H_x$ = 100 Oe. **(C)** Simultaneous measurement of the relative change in the Hall signal (black) together with the measurement of the magnetization change by direct MOKE imaging (red) for the sequence of images shown in (B). The change in magnetization corresponds to the red dashed area indicated in image #8. The change in the Hall resistance is due to $R_H^{AHE,SMR}$ (see Section S1 and Fig. S2), which reflects the change of m$_z$. From these measurements we infer that the electrical reading of the magnetization state of TmIG is only sensitive to changes in the magnetization in the vicinity of the Hall cross. Comparison of the Hall and MOKE data reveals that the sensitivity of the Hall measurement extends slightly away from the Hall cross. Spreading of domains beyond the Pt current line can also influence $R_H^{AHE,SMR}$. Electrical measurements alone do not allow to determine the DW position and displacement with precision after the application of current pulses.

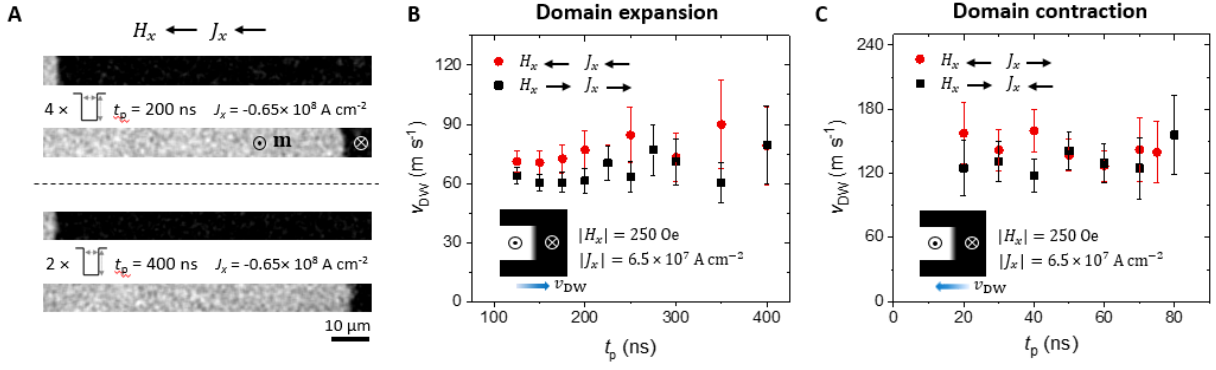

**Fig. S6. Dependence of $v_{\text{DW}}$ on the pulse length.** (**A**) Differential MOKE images showing the displacement of an up-down DW after the application of $N = 4$ current pulses of $t_p = 200$ ns and $N = 2$ pulses of $t_p = 400$ ns. $H_x = -250$ Oe and $J_x = -6.5 \times 10^7$ A cm$^{-2}$ in both cases. The same displacement $\Delta x$ is observed for both sequences of current pulses, leading to the same estimate for the DW speed $v_{\text{DW}} = \Delta x / (N_p\, t_p)$ (see Methods). (**B**) Systematic evaluation of $v_{\text{DW}}$ for an up-down DW during domain expansion as a function of $t_p$ in the regime where DWs move upon the application of current pulses but are not strongly influenced by Joule heating. (**C**) Same as (B) for domain contraction. Within the experimental error, $v_{\text{DW}}$ remains roughly constant although $t_p$ changes by a factor of ~4, which allows us to conclude that the DW speed is accurately evaluated upon changing $t_p$ and not influenced by DW inertia. We attribute the slight increase of $v_{\text{DW}}$ at large $t_p$ to Joule heating. For the measurements presented in Fig. 4 we reduced the influence of Joule heating by measuring $v_{\text{DW}}$ at the lower edge of the parameter space $t_p$ at any given $H_x, J_x$ combination. Joule heating (as well as the Oersted field, see Section S2) is found to influence the DW dynamics and magnetization switching at current densities $\gtrsim 1 \times 10^8$ A cm$^{-2}$.

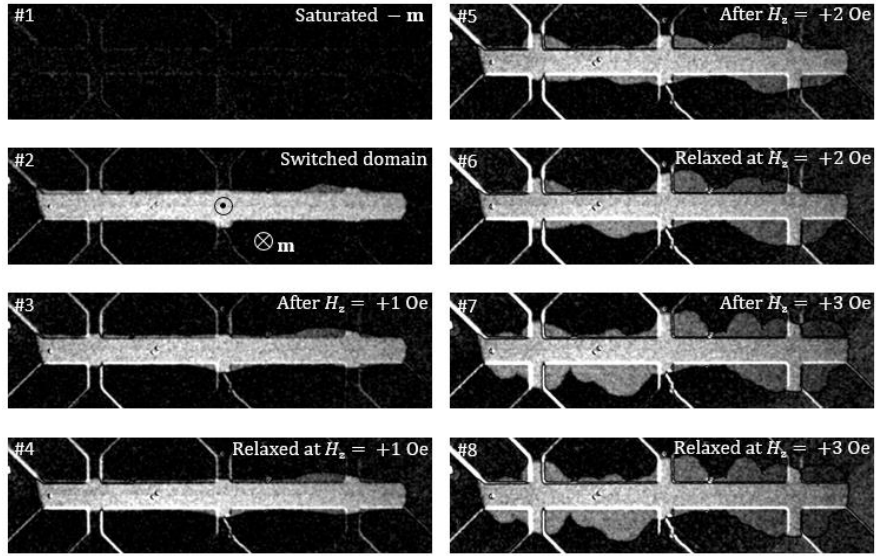

**Fig S7. Evaluation of the domain wall depinning field.** Sequence of differential MOKE images. #1, The TmIG film is initially satured **−m** by applying $H_z = -100$ Oe (dark contrast). #2, The magnetization underneath the Pt current line is switched to **+m** (bright contrast) by applying a forward switching current pulse. #3 to #8, An out-of-plane field $+H_z$ is applied and the evolution of the DWs of the **+m** domain tracked with time and strength of the magnetic field. Images #3/5/7 are taken right after applying $H_z = +1/2/3$ Oe, whereas images #4/6/8 are the subsequent MOKE images taken after waiting 20 sec at the same field. A first tiny DW displacement is already observed for $H_z = +1$ Oe. At $H_z = +2$ Oe the DW is clearly depinned and the domain extends significantly through the TmIG film. We thus determine that the DW depinning field in TmIG is on the order of ∼1-2 Oe.

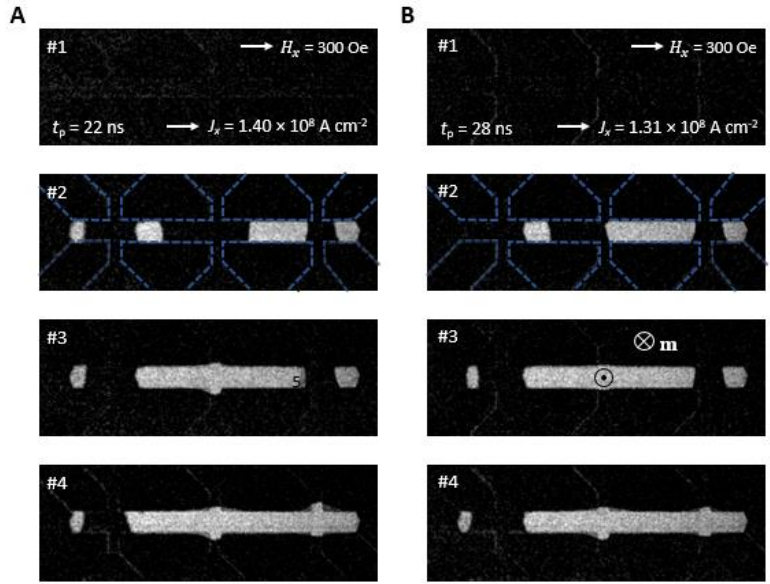

**Fig. S8. Multiple domain nucleation at large current densities.** (**A** and **B**) Examples of multiple nucleation of domains at large current densities. Images #1 to #4 correspond to a series of forward switching current pulses. In contrast to low current densities, where nucleation takes place at a defect site (see Fig. 3B and Fig. S3), here the nucleation takes place preferentially in the region comprised between two adjacent Hall crosses (not right at the cross, see #2), where the current density is larger.

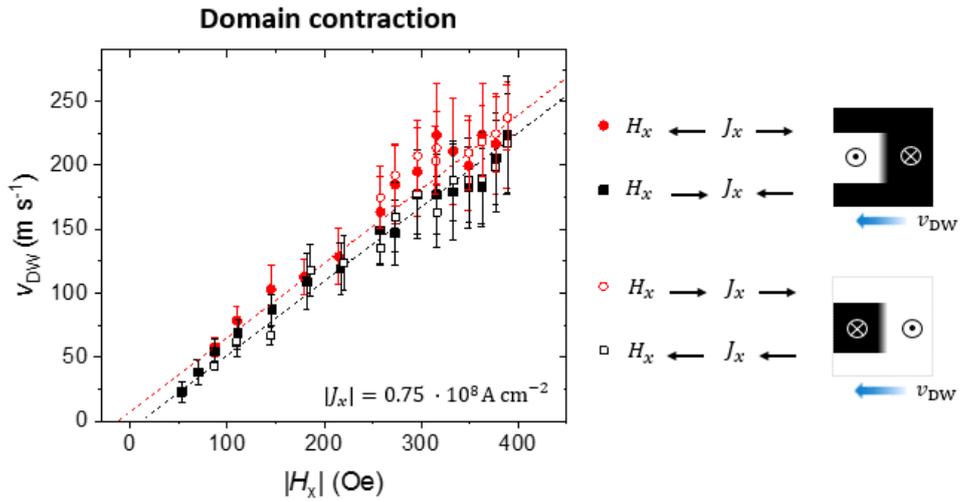

**Fig. S9. Comparison of $v_{DW}$ between an up-down and a down-up domain wall.** $v_{DW}$ as a function of $H_x$ for a down-up DW, and for the two combinations of $H_x$ and $J_x$ that lead to a domain contraction (open symbols), together with the data presented in Fig. 4D for a contracting up-down DW (solid symbols). Within the experimental error, both DWs exhibit a similar larger/slower speed when moving against/along $J_x$, confirming the weak negative DMI found in TmIG/Pt. See also Fig. S10, where the DW structure and current-induced DW motion is discussed for both up-down and down-up DWs and for all four combinations of $H_x$, $J_x$ directions.

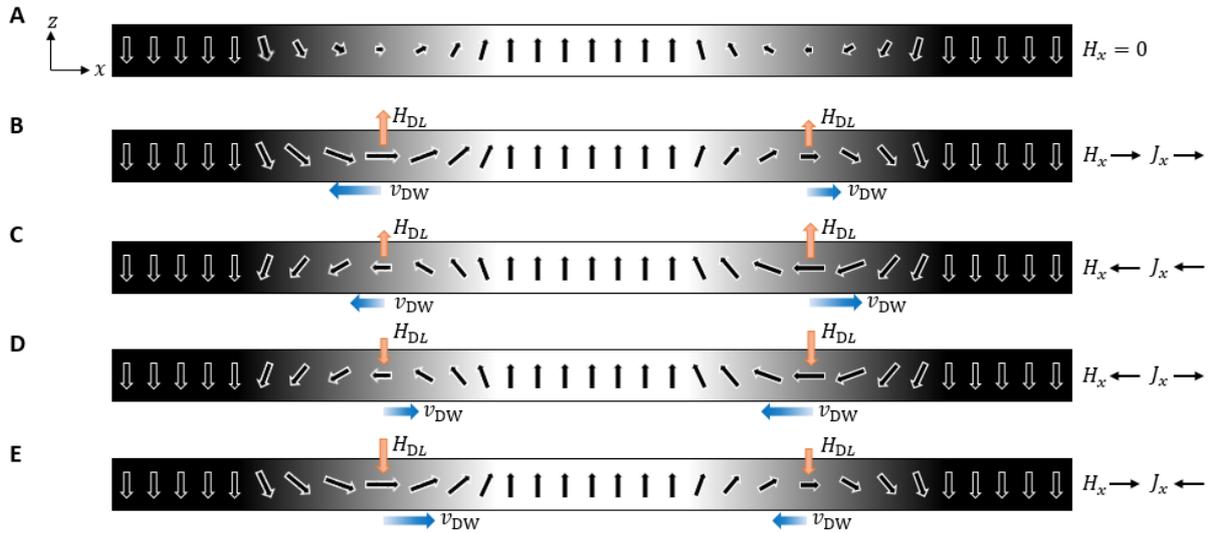

**Fig. S10. Schematics of the domain wall structure and current-induced domain wall motion in the presence of $H_x$ and DMI.** (**A**) Cross-sectional schematics (*xz*-plane) of a down-up and an up-down DW (positioned along the y-direction) in TmIG/Pt illustrating its internal structure in the absence of $H_x$. The finite left-handed Néel chirality of the DWs reflects the negative DMI found in TmIG/Pt (see Fig. 2 and Fig. 4, C and D). (**B** to **E**) The internal magnetization of the DWs can be controlled by applying a magnetic field $H_x$ exceeding the effective DMI field $H_{DMI}$, where the latter is positive (negative) for a down-up (up-down) DW (A). Consequently, upon the application of $H_x$, the internal magnetization of up-down and down-up DWs has either a larger or smaller projection on the direction of $J_x$. As the amplitude of the damping-like SOT is maximal when the DW magnetization is parallel or antiparallel to $J_x$, up-down and down-up DWs move at different speed. The damping-like SOT is represented by the corresponding effective field $H_{DL}$ in the figure. Application of current pulses drive the DWs as indicated for all four possible combinations of $H_x$ and $J_x$ applied (see B to E). DWs are driven at larger (smaller) velocities when moving against (towards) $J_x$, which is consistent with the results presented in Fig. 4 and Fig. S9.

## Section S3. Scanning NV magnetometry

**Definition of the polar angles $\phi_{NV}$ and $\theta_{NV}$**

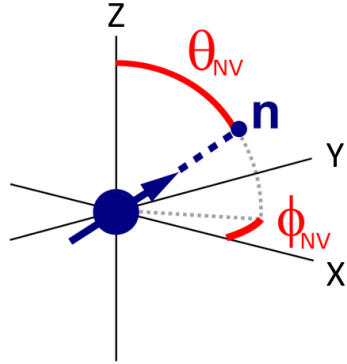

**Fig. S11.** Definition of the polar angles $\phi_{NV}$ and $\theta_{NV}$ with respect to the laboratory coordinate frame (see Fig. 2). $\theta_{NV}$ = (55 ± 2)° is determined by sample fabrication. Analysis of the stray field $B_{NV}$ allows us to determine that $\phi_{NV}$ = (83 ± 3)°.

**Determination of the stand-off distance *d*:**

The sample-to-sensor distance *d* is inferred *in situ* from the change in the stray field $B_{NV}$ at the TIG/Pt edge (see Fig. 2B) due to the different magnetic response of TmIG relative to TmIG/Pt (see Fig. 2C), which we mainly attribute to a proximity induced polarization of the Pt layer (see second dataset taken in a reference TmIG film). The change in magnetization gives rise to a stray field that is added to the field produced by the DW. In order to disentangle both fields we use a differential approach by shifting the images along the y-axis and subtracting them from each other, see Fig. S12. In this way we cancel the domain wall stray field and end up with a stray field showing only two Pt edges. The magnetic stray field emanating from the Pt edges, with the image orientated such that they are aligned with the *y*-direction, can be described analytically as

$$B_x = \frac{\mu_0 M_{S,Pt} t}{2\pi}\left[\frac{d-x_1}{d^2+(x-x_1)^2} - \frac{d-x_0}{d^2+(x-x_0)^2}\right]$$

$$B_y = 0$$

$$B_z = -\frac{\mu_0 M_{S,Pt}\cdot t}{2\pi}\left[\frac{x-x_1}{d^2+(x-x_1)^2} - \frac{x-x_0}{d^2+(x-x_0)^2}\right], \quad (S5)$$

where $x_0, x_1$ denote the edge positions and $M_{S,Pt} t$ is the surface magnetization of Pt. By fitting this model to the experimental line scan we can extract a value for the stand-off distance *d* and $M_{S,Pt}$. Further fit parameters are the edge positions and the azimuth angle of the NV vector orientation. To improve statistics and estimate the error, we fit several line scans across the stripe using the same starting parameters and extract the mean value and standard error. For the image shown in Fig. 2B and Fig. S12, we obtain $d = 104 \pm 3.3$ nm. Further stand-off distance measurements on a separate calibration sample (Pt/Co/AlOx) before and after the domain wall measurement are in good agreement with this value. To compensate for the fact that we have not included the height profile of the stripe as well as the Dzyaloshinskii-Moriya interaction, we deliberately over-estimated the error in the stand-off distance to be larger than the fit errors and set it to 5 nm.

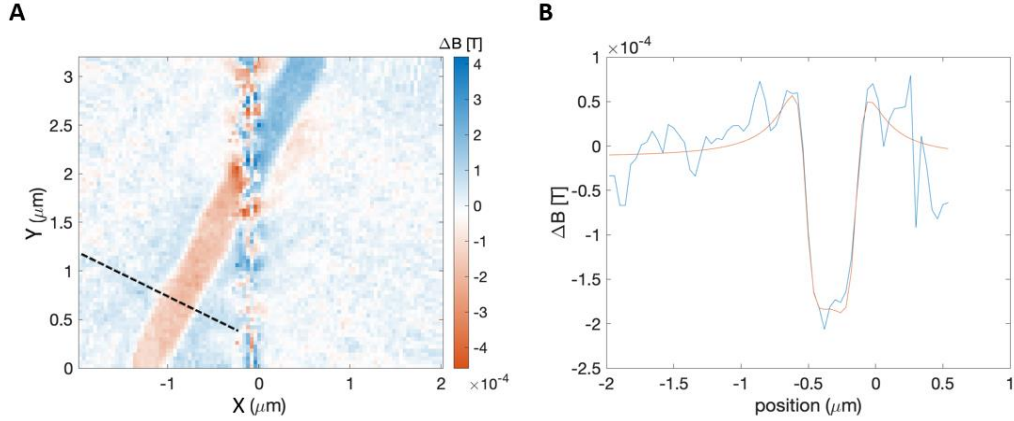

**Fig. S12. In situ calibration of the NV stand-off distance on TmIG/Pt.** (**A**) Differential stray field image. The plotted image was obtained by subtracting two original images (Fig. 2B) that were shifted relative to each other by 800 nm along the Pt edge. (**B**) Representative line cut of the differential image along the dashed line indicated in (A), together with the fit according to Eq. (S5).

**Fits of the domain wall magnetization profile:**

We fit several line scans of the DW in TmIG and TmIG/Pt using the same starting fit parameters as described in the Methods section. The resulting fit parameters $p_i$ are averaged, with weights according to their variance. The sum is taken over all line scans corresponding to the selected region. The nominal value of the parameters are calculated as $\bar{p}_i = \sum_{k=1}^{N} w \cdot p_i$. The standard deviation of each parameter is then $\sigma_i = \sqrt{\sum_{k=1}^{N} w \cdot p_i^2 - \bar{p}_i^2 + \left(\delta d \frac{\partial \bar{p}_i}{\partial \bar{d}}\right)^2}$. The last term describes the uncertainty due to the error of the stand-off distance $\delta d$. Since the fitting algorithm assumes the stand-off distance to be fixed, it is not included in the fitting results directly. We estimate $\frac{\partial \bar{p}_i}{\partial \bar{d}}$ via finite differences by repeating the fitting procedure with the stand-off distance equal to the error boundaries $d^{\pm} = \bar{d} \pm \delta d$ with $\bar{d} = 104$ nm.

**Second dataset: reference TmIG film**

We recorded a second dataset with a different scanning probe ($d = 66 \pm 6$ nm) on an unprocessed TmIG film of thickness 8.5 nm. Analysis of this second dataset yielded a chiral angle of $\psi = (180 \pm 0)°$, which corresponds to a pure left-handed chiral Néel domain wall structure, a DW width of $\Delta_{DW} = (20 \pm 4)$, and an out-of-plane magnetization $M_z = (53.3 \pm 3.5)$ kA m$^{-1}$. These values are consistent with the ones presented in Fig. 2 for TmIG(8.3 nm), supporting the finding of a strong DMI in an all-oxide structure without the presence of a metallic heavy metal layer.

**Supplementary References**